\documentclass{article}
\usepackage{graphics}

\input{lhk4.mac}        

\input{lhk4.pre}        
\input{lhk4.fig}        

\input{1up.inp}

\begin{document}


\begin{titlepage}

\begin{tabular}{l}
\noindent\DATE
\end{tabular}
\hfill
\begin{tabular}{l}
\PPrtNo
\end{tabular}

\vspace{1cm}

\begin{center}
\renewcommand{\thefootnote}{\fnsymbol{footnote}}
{
\LARGE \TITLE
}

\vspace{1.25cm}
{\large  \AUTHORS}

\vspace{1.25cm}

\INST
\end{center}

\vfill

\ABSTRACT                 

\vfill

\newpage
\end{titlepage}


\renewcommand{\thefootnote}{\alph{footnote}}


\section{Introduction}

\label{sec:Intro}

Recent measurements of charm production in deep inelastic scattering (DIS)
at HERA \cite{H1charm96,ZeusCharm} have shown that up to 25\% of the total
cross-section at small-x contains charm in the final state. This is within
the expectations of perturbative QCD based on conventional parton
distributions. We are now in a position to utilize this process to study
details of the production mechanism of heavy quarks in general, and to
extract useful information on the charm and gluon structure of the proton in
particular.

Conventional perturbative QCD (PQCD) theory is formulated in terms of
zero-mass quark-partons. For processes depending on one hard scale $Q,$ the
well-known factorization theorem then provides a straightforward procedure
for order-by-order perturbative calculations, as well as an associated
intuitive parton picture interpretation of the perturbation series. Heavy
quark production presents a challenge in PQCD because the heavy quark mass, $%
m{H}$ $(H=c,b,t),$ provides an additional hard scale which complicates the
situation -- it requires a different organization of the perturbative series
depending on the relative magnitudes of $m{H}$ and $Q$.

The two standard methods for PQCD calculation of heavy quark processes
represent two diametrically opposite ways of reducing the two-scale problem to
a one-scale problem. (i) In the conventional \emph{parton model approach} used
in many global QCD analyses of parton distributions \cite {EHLQ,MRS,CTEQ} and
Monte Carlo programs, the zero-mass parton approximation is applied to a heavy
quark calculation
as soon as the typical energy scale\footnote{%
We use $Q$ as the generic name for a typical kinematic physical scale. It could
be $Q$, $W$, or $p{T}$, depending on the process.} of the physical process $Q$
is above the mass threshold $m{H}$. This leaves $Q$ as the only apparent hard
scale in the problem. (ii) In the \emph{heavy quark approach} which played a
dominant role in ``NLO calculations'' of the production of heavy quarks
\cite{NDE,SvNorg,FMNR97}, the quark $H$ is always treated as a ``heavy''
particle and never as a parton. The mass parameter $m{H}$ is explicitly kept
along with $Q$ as if they are of the same order, irrespective of their real
relative magnitudes.

The co-existence of these two opposite approaches represents an uneasy
dichotomy in the current literature. On physical grounds, the zero-mass parton
picture of heavy quarks should be applicable at energy scales very much larger
than the relevant quark mass, $m{H}$ $\ll $ $Q$, whereas the heavy quark
approach (often referred to as the fixed-flavor-number (FFN) scheme) should be
more appropriate at energy scales comparable to the quark mass $m{H}$ $\sim $
$Q$. The actual experimental regime often lies \emph{in between} these two
extreme regions, where the validity of either approach can be called into
question. There is, however, a natural way to incorporate both approaches in a
unified framework in PQCD which provides a smooth transition between the two.
This has been formulated in a series of papers over the years,
Refs.\ \cite{ColTun,OlnTun87,AOT94a,ACOT,Collins97}%
\footnote{%
See Ref.~\cite{dis97tung} for a brief review.}, %
which has now been adopted, in different guises, by most recent literature on
heavy quark production in PQCD. \cite{RobertsT,BMSvN,NasonEtAl,CSvN}

To see the basic ideas behind this unified picture, let us focus explicitly on
the production of charm ($H=c$) in deep inelastic scattering. All
considerations
apply to a generic heavy quark. Consider the PQCD calculation of the $%
F{2}(x,Q)$ structure function which receives substantial contribution from
charm production as mentioned earlier. The underlying physical ideas are
illustrated graphically in Fig.\ \ref{fig:cartoon} where the charm contribution
to this structure function, denoted $F{2}^{c}(x,Q)$, is
plotted as a function of $Q$ at some fixed value of $x$.%
\figcartoon%
Near threshold $Q\sim m{c}$, it is natural to consider the charm quark as a
heavy particle, and to adopt the 3 active-parton-flavor scheme of calculation
(the ``heavy quark approach''). As $Q$ becomes large compared to $m{c}$, the
fixed-order calculation in this approach becomes unreliable since the
perturbative expansion contains terms of the form $\alpha {s}^{n}\log
^{n}\left( m{c}^{2}/Q^{2}\right) $ at any order $n$, which ruin the
convergence of the series---these terms are \emph{not infra-red safe} as
$m{c}$ $\rightarrow 0$ or $Q\rightarrow \infty $.
Thus the uncertainty of the 3-flavor calculation grows as $Q/m{c}$ becomes
large. This is illustrated in Fig.\ \ref{fig:cartoon} as an error
band marked by horizontal hashes which is narrow near threshold but becomes
ever wider as $Q/m{c}$ increases.
On the other hand, starting from the high energy end
($Q\gg m{c}$), the most natural calculational scheme to adopt is the
conventional 4-flavor scheme with active charm partons. (In this approach, the
infra-red unsafe large logarithms mentioned earlier are ``resummed" and
absorbed into the finite charm parton distributions.) However, as we go down
the energy scale toward the charm production threshold region, the 4-flavor
calculation becomes unreliable because the approximation $m{c}=0$ deteriorates
as $Q\rightarrow m{c}$. The uncertainty band of such a calculation is outlined
in Fig.\ \ref{fig:cartoon} by the vertical hashes -- it is narrow at high
energies, but becomes increasingly wider as one approaches the threshold region.

The intuitive ideas embodied in Fig\ \ref{fig:cartoon} illustrate that: (i)
these two conventional approaches are individually unsatisfactory over the full
energy range, but are mutually complementary; and (ii) the most reliable PQCD
prediction for the physical $F{2}(x,Q)$, at a given order of calculation, can
be obtained by utilizing the most appropriate scheme at that energy scale $Q$,
resulting in a composite scheme, as represented by the cross-hashed region in
Fig.\ \ref{fig:cartoon}.
The use of a composite scheme consisting of different numbers of flavors in
different energy ranges, rather than a fixed number of flavors, is familiar in
the conventional zero-mass parton picture.  The new formalism espoused in
Refs.\
\cite{ColTun,OlnTun87,AOT94a,ACOT} provides a quantum field
theoretical basis \cite{CWZ,Collins97}
for this intuitive picture in the presence of non-zero quark mass.

This generalization brings
about several important distinguishing features and insights. First, after the
hard cross-section is rendered infra-red safe by factorizing out the $\ln
(m{c})$ terms, the remaining finite dependence on $m{c}$ can be kept in the
hard cross-sections to
maintain better accuracy in the intermediate energy region.%
\footnote{%
By contrast, in the standard literature the $m{c}\rightarrow 0$ limit is tied
to the proof of factorization \cite{ZeroMfact} in the first place. This
association is not needed, cf.\ \cite{Collins97}}%
One can then show that the
unified formalism reproduces the two conventional
approaches in well-defined ways in their respective regions of applicability
\cite{ColTun}-\cite{Collins97}.
Secondly, it has become clear recently that there is much inherent flexibility in the
choice of the transition energy scale (cf.\ Fig.~\ref{fig:cartoon}) as well as
the detailed matching condition between the 3- and 4-flavor calculations.
This
makes it possible to have a variety of different implementations of the new
formalism \cite{RobertsT,BMSvN,NasonEtAl,CSvN}, with different emphases. Properly
understood, this feature can be exploited to lend strength to the formalism;
but, during the development of the theory, the subtleties have given rise to
much misunderstanding and confusion in existing literature.

In this paper, we study charm production in DIS using the general mass
formalism, and compare the results with recent HERA data. The main goals of
this work are:

\begin{Simlis}[]{0em}
\item%
(i) Through this specific example, we give a concise and careful presentation
of the general formalism, including the relatively simple operational procedure
for calculating its various components at any order of $\alpha_s$.
We hope this will help fill the gap
between the original, relatively sketchy, ACOT paper \cite{ACOT} and the
recent, more technical, all-order proof of the formalism by Collins
\cite{Collins97}.
\item%
(ii) We carry out the numerical calculations to make concrete the intuitive
ideas illustrated in Fig.~\ref{fig:cartoon}; and to demonstrate the validity of
the physical principles underlying the composite scheme.
\item%
(iii) We show that the flexibility of the general formalism mentioned above can
result in efficient PQCD calculations of inclusive quantities at relatively low
order in $\alpha_s$ compared to FFN calculations -- because the relevant
physics has been effectively ``resummed'' by the appropriate scheme adopted for
the given energy scale.

\end{Simlis}

Item (i) is basic; it forms the foundation for the other two parts. However,
since this part is about the clarification of the existing theoretical
formalism rather than new work, and since not all readers are concerned with
theoretical precision, we have elected to place it in the Appendix. Hopefully, this will
increase the accessibility of the main body of the paper. In regards to the
underlying physics, we believe the discussion in the appendix should make a useful
contribution to clear the confusion and misunderstanding among the various
approaches that have been proposed in the recent literature following
\cite{ACOT}.

Based on the terminologies discussed in this Appendix, in Sec.~\ref{sec:TotInc}
we describe the complete order $\alpha_s$ calculation carried out in this paper
in relation to previous work on this subject. The new calculation extends the
validity of the original ACOT results to NLO in the high energy regime -- on
the same level as the conventional zero-mass total inclusive structure
functions. In addition, the new perspective, as illustrated in Fig.~\ref{fig:cartoon}
above and discussed quantitatively in the paper proper, allows a re-assessment
of the physical predictions of the order $\alpha_s$ calculation near the
threshold region, making it a viable alternative to the order $\alpha_s^2$ FFN
calculation. In Sec.~\ref{sec:inclusive} we present the numerical results on
inclusive charm production, and demonstrate that the validity and the
efficiency of the general formalism, as described in (ii) and (iii) above, are
indeed seen at this order. We show that very good agreement with recent HERA
data on inclusive $F{2}^{c}$ is obtained in practice. For this study, we have
developed a new implementation of the generalized \msbar\ formalism using Monte Carlo
methods. This implementation allows the computation of differential
distributions, with kinematic cuts, such as $d\sigma ^{D^{\ast
}}/dp{t},\;d\sigma ^{D^{\ast }}/dQ,$ and $d\sigma ^{D^{\ast }}/d\eta $ which
we present in Sec.~\ref{sec:tagged}. These results, are in qualitative
agreement with available data from HERA.  However, they are not as good as
those of the order $\alpha {s}^{2}$ calculation in the 3-flavor scheme. This
is to be expected for differential distributions with experimental cuts, since
the (resummed) low-order calculation contains more severe approximations to the
kinematics of the final state partons. For future quantitative studies, the
general formalism needs to be expanded to incorporate higher-order results
(adaptable from existing FFN calculations). This point is discussed in the
concluding section, along with other observations.

                      


\section{Total Inclusive Structure Functions in the general formalism}

\label{sec:TotInc}

We consider the inclusive DIS structure functions, such as $F_{2}$, focusing
on the contribution of a massive quark. For definiteness, we assume that the
only relevant quark with non-zero mass is the charm quark. The generic
leptoproduction process is depicted in Fig.~\ref{fig:lhProc}: 
\begin{equation}
\ell _{1}+A\longrightarrow \ell _{2}+X,  \label{eq:LepProc}
\end{equation}
where $A$ is a hadron, $\ell _{1,2}$ are leptons, and $X$ represents the
summed-over final state hadronic particles. Note that $X$ may or may not
contain a visible heavy-flavor hadron. \figlhProc After the calculable
leptonic part of the cross-section has been factored out, we work with the
hadronic process induced by the virtual vector boson $\gamma ^{\ast }$ of
momentum $q$ and polarization $\lambda $: 
\begin{equation}
\gamma ^{\ast }(q,\lambda )+A(P)\longrightarrow X(P_{X}).
\label{eq:BosonProc}
\end{equation}
Although our considerations apply to DIS processes induced by $W$ and $Z$ as
well, we shall explicitly refer to the neutral current interaction with the
exchange of a virtual photon $\gamma ^{\ast }$ in order to be concrete. The
cross-section is expressed in terms of the hadronic tensor 
\begin{equation}
W_{\lambda \sigma }(q,P)=\frac{1}{4\pi }\mathrel{\mathop{\overline{\sum
}}\limits_{X(P_{X}),spin}}\langle P|e_{\lambda }^{\ast }\cdot J^{\dagger
}|P_{X}\rangle (2\pi )^{4}{\delta ^{(4)}\left( P+q-P_{X}\right) }\langle
P_{X}|e_{\sigma }\cdot J|P\rangle .  \label{eq:HadTensor}
\end{equation}
where $\overline{{\sum }}$ denotes a sum over all final hadronic states. In
most cases, it suffices to consider the diagonal elements of the tensor $%
F^{\lambda }\equiv W_{\lambda \lambda }(q,P).$

The factorization theorem in the presence of non-zero quark masses --
assumed in \cite{ACOT} and established to all orders in PQCD \cite{Collins97}
-- states that the inclusive cross-section can be written as a convolution: 
\begin{equation}
F_{A}^{\lambda }(Q^{2},x,..)=\stackunder{a}{\sum }f_{A}^{a}(x,\mu )\otimes 
\widehat{\omega }_{a,\lambda }(x,Q/\mu ,Q/m_{c},\alpha _{s}\left( \mu
\right) )\;+\;\mathcal{O}\Bigl(\Lambda /Q\Bigr)^{p}  \label{eq:HadXsec0}
\end{equation}
where $f_{A}^{a}$ is the distribution of parton $a$ inside the hadron $A$, $%
\widehat{\omega }_{a,\lambda }$ is the perturbatively calculable hard
cross-section for $\gamma ^{\ast }+a\rightarrow X$, $p$ is some positive
number, $\mu$ denotes collectively the renormalization and factorization
scales, and a convolution in the $x$ variable is implied. The helicity
structure functions $F^{\lambda }$ are simply related to the familiar $%
F_{1,2,3}$ \cite{AOT94a}.

The exact way that the physical structure function factorizes into the
long-distance ($f_{A}^{a}$) and the short-distance ($\widehat{\omega }%
_{a,\lambda }$) pieces on the right-hand side of Eq.~(\ref{eq:HadXsec0})
depends on the scheme used to define the parton distributions. The physical
structure function $F_A^\lambda$ should be independent of any calculational
scheme; therefore, the definition of the hard cross-sections is determined
by the subtraction procedure used to define the parton distributions. As
discussed in the Introduction, the general formalism consists of the
3-flavor scheme at low energy scales, the 4-flavor scheme at high energy
scales, and a suitably chosen transition region where matching conditions
between the two schemes are applied. The precise description of these
elements of the formalism is given in the Appendix. Operational definitions
of quantities needed in subsequent discussions are also discussed in some
detail there.

\subsection{Previous calculations}

\label{sec:PrevCal}

To put the current calculation in context, we first summarize the existing
calculations of leptoproduction of charm using the precise definitions given
in the Appendix.

\begin{Simlis}{1em}
\item  \textbf{NLO 3-flavor }(3$\alpha _{s}^{2}$) \textbf{calculation}
\textbf{\cite{SvNorg}}: Most dedicated calculations of heavy flavor production
in recent years have been carried out in this scheme. The LO process is
$\mathcal{O}(\alpha _{s}^{1})$ heavy-flavor creation (HC), $\gamma ^{\ast
}g\rightarrow c \bar{c}$. The NLO processes consist of the $\mathcal{O}(\alpha
_{s}^{2})$ virtual corrections to $\gamma ^{\ast }g\rightarrow c\bar{c}$ as
well as the real HC $\gamma ^{\ast }l\rightarrow c\bar{c}l$ process, where $l$
denotes any light parton. Cf.\ Fig.~\ref{fig:Threefl}. This calculation becomes
questionable when $Q\gg m_{c}$ -- {\it i.e.,} it ceases to be ``NLO''
in accuracy, as indicated in Fig.\ \ref{fig:cartoon},
because the perturbative expansion is
actually in $\alpha_s ln(Q^2/ m_{c}^2)$ for large $Q$.
\figThreefl

\item  \textbf{Zero-mass 4-flavor} (ZM) (4$\alpha _{s}^{1};$ $m_{c}=0$) \textbf{%
calculation:} This is the formalism used in most conventional QCD parton model
calculations and popular Monte Carlo programs. It represents an approximation
to the general mass (GM) 4-flavor scheme by setting $m_{c}=0$
in the hard cross-sections $\hat{\omega}_{a}(x,\frac{Q}{\mu },\frac{m_{c}}{Q}%
,\mu )\stackrel{m_{c}/Q\rightarrow 0}{\longrightarrow }\hat{\omega}%
_{a}^{m_{c}=0}(x,\frac{Q}{\mu },\mu ).$ The LO contribution consists of the
$\mathcal{O}(\alpha _{s}^{0})$ $\gamma ^{\ast }c\rightarrow c$ heavy-flavor
excitation (HE) process. The NLO contribution consists of the
$\mathcal{O}(\alpha _{s}^{1})$ virtual corrections to $\gamma^{\ast
}c\rightarrow c$ plus the real HC $\gamma ^{\ast }g\rightarrow c\bar{c}$ and,
$\gamma ^{\ast }c\rightarrow gc$ HE processes. Cf.\ Fig.~\ref{fig:Fourfl}. This
calculation is unreliable in the threshold region, as mentioned in the
introduction.

\figFourfl

\item  \textbf{LO generalized \msbar\ calculation} (ACOT) \cite{ACOT}:
This represents the simplest implementation of the generalized formalism
\cite{ACOT}. It emphasizes the overlapping physics underlying the 3-flavor and
4-flavor calculations in the region not far above threshold. The 4-flavor
calculation consists of HE $\gamma^{\ast}c\rightarrow c$
(Fig.~\ref{fig:Fourfl}a) plus $m_c \neq 0$ HC $\gamma ^{\ast}g\rightarrow
c\bar{c}$ (Fig.~\ref{fig:Fourfl}d), with the requisite subtraction term which
removes the $m_{c}$ mass-logarithm. Mathematically, the result on
$F_2(x,Q,\mu)$ can be shown to match that of the $\mathcal{O}(\alpha _{s}^{1})$
3-flavor scheme calculation (HC $\gamma ^{\ast }g\rightarrow c \bar{c}$) as the
(unphysical) factorization scale $\mu$ approaches $m_{c}$ from above
\cite{ACOT}. Physically, the predicted behavior of $F_2^c$ for $Q \sim m_c$
will depend on the choice of $\mu$ as a function of the physical variables,
cf.\ next section. At a high energy scale, $\mu \sim Q\gg m_{c}$, this
calculation only approximates the 4-flavor NLO results: it contains the most
important $\mathcal{O}(\alpha _{s}^{1})$ term, HC $\gamma ^{\ast }g\rightarrow
c\bar{c}$ (because of the large gluon distribution and the need for matching),
but does not include the smaller $\mathcal{O}(\alpha _{s}^{1})$ terms
represented by Fig.~\ref{fig:Fourfl}b,c. This calculation has been further
studied by Kretzer and Schienbein \cite{KretzerS} and Kr\"amer {\it et al}
\cite{KOS}.

\item \textbf{Variations on the ``variable flavor number'' theme:}
In recent years, the general approach proposed in \cite{ColTun,ACOT} has been
adopted by other groups, starting from different historical perspectives,
\cite{RobertsT,BMSvN,NasonEtAl,CSvN}. In contrast to the fixed flavor-number
(FFN) approach, these are usually referred to as being in the variable
flavor-number (VFN) scheme. This terminology has caused some confusion, since
although the common theme is that of \cite{ColTun,ACOT}, as shown in Fig.\
\ref{fig:cartoon}, the various implementations differ considerably. Whether a
particular implementation is self-consistent within the general framework of
PQCD, or whether two implementations are compatible within the accuracy of the
perturbative expansion, are often unclear or controversial (e.g.\
\cite{ColMarRys}) because of the complexity of the multi-scale problem and
because of possible misunderstandings. It is beyond the scope of this paper to
critically review these approaches. By pursuing the goals described in the
introduction, we hope to provide a clearer picture of the general formalism of
\cite{Collins97}, hence a better basis to help address some of the
controversial issues in the future.
\end{Simlis}

\subsection{The full order $\protect\alpha_s$ generalized \msbar\ calculation%
}

The calculation of charm contribution to the inclusive structure functions
reported in this paper completes the order $\alpha_{s}$ calculation in the
general formalism, initiated in \cite{ACOT}, including all the hard
processes of Fig.~\ref{fig:Fourfl}. The additional terms, although
relatively small numerically at current energies, are required to make the
4-flavor part of this calculation truly NLO at high energies, so that it
becomes equivalent to the conventional zero-mass NLO theory used in most
modern analyses of precision DIS data. In the remainder of this section we
discuss the theoretical issues and uncertainties in the order $\alpha_s$
4-flavor calculation at all energy scales, in order to address the issues
highlighted at the end of the introduction, Sec.~\ref{sec:Intro}.

At order $\alpha _{s}^{1}$ the 3-flavor component of the composite scheme
consists of only the (unsubtracted) $\mathcal{O}(\alpha _{s}^{1})$ HC $%
\gamma ^{\ast }g\rightarrow c\bar{c}$ process. The result is standard.
Therefore, our main calculation concerns the \emph{4-flavor scheme component}%
. At order $\alpha _{s}^{1}$ the right-hand side of Eq.~(\ref{eq:HadXsec0})
consists of three terms 
\begin{equation}
\begin{array}{lll}
F^{(4)}_{A,\lambda }(Q^{2},x,m_c,\mu) & = & \;\;\;f_{A}^{c}\otimes \,^{0}%
\widehat{\omega }_{c,\lambda }^{c} \\ 
&  & +\;f_{A}^{g}\otimes \,^{1}\widehat{\omega }_{g,\lambda }^{c\bar{c}} \\ 
&  & +\;f_{A}^{c}\otimes \,^{1}\widehat{\omega }_{c,\lambda }^{cX} \\ 
&  & +\;\text{light-parton\ contributions}\ ,
\end{array}
\label{eq:HadXsec4fl0}
\end{equation}
where the superscript (4) indicates that this is a 4-flavor calculation, and
the \emph{hard-scattering cross-sections} $^{i}\widehat{\omega}_{a,\lambda
}^{X}$ for the various subprocesses are calculated from the corresponding 
\emph{partonic cross-sections} $^i\omega_{a,\lambda}^X$ (without the hat)
according to the procedures described in the Appendix. A description of each of the terms follows:

\begin{Simlis}{1em}
\item  The leading order $\gamma ^{*}c\rightarrow c$ (Fig.~\ref{fig:Fourfl}a)
partonic cross-section ${}^0\omega _{c,\lambda }^c$ is
infra-red safe, thus
\begin{equation}
^0\widehat{\omega }_{c,\lambda }^c=\,^0\omega _{c,\lambda }^c\ .
\label{eq:Hardxsc0}
\end{equation}

\item  The $\gamma ^{*}g\rightarrow c\bar{c}$ (Fig.~\ref{fig:Fourfl}d)
partonic cross-section contains a single power of $\ln
\mu ^2/m_c^2$ which can be factorized into the charm distribution function by
the subtraction \cite{ACOT}
\begin{equation}
^1\widehat{\omega }_{g,\lambda }^{c\bar{c}}=\,^1\omega _{g,\lambda }^{c\bar{c%
}}-\,^1\tilde{f}_g^c\otimes \,^0\omega _{c,\lambda }^c\ ,
\label{eq:Hardxsc1}
\end{equation}
where the cancelling logarithm with mass-singularity resides in the
$\mathcal{O}(\alpha _s)$ perturbative parton distribution function
\begin{equation}
^1\tilde{f}_g^c=(\alpha _s/2\pi )P_{g\rightarrow q}(x)\ln (\mu ^2/m_c^2)
\label{eq:fgc1}
\end{equation}

\item  The virtual correction to $\gamma ^{*}c\rightarrow c$
(Fig.~\ref{fig:Fourfl}b) plus the real $\gamma ^{*}c\rightarrow gc$
(Fig.~\ref{fig:Fourfl}c) partonic process also contain $\ln \left( \mu
^2/m_c^2\right) $ terms which are factorized into the charm distribution
function by the subtraction
\begin{equation}
^1\widehat{\omega }_{c,\lambda }^{cX}=\,^1\omega _{c,\lambda }^{cX}-\,^1%
\tilde{f}_c^c\otimes \,^0\omega _{c,\lambda }^c\ ,  \label{eq:Hardxsc2}
\end{equation}
where the logarithm appears in the $\mathcal{O}(\alpha _s^1)$ perturbative
parton distribution function%
\footnote{%
We have calculated these terms keeping a finite charm quark mass. As
discussed in Ref.~\cite{Collins97,KOS}, it would also have been consistent to
calculate diagrams with an initial state charm quark using a zero charm
quark mass. The errors near threshold in both methods of calculation are
comparable and of order $\alpha_s^2$.}%
\begin{equation}
^1\tilde{f}_c^c={\frac{\alpha _s}{2\pi }}{\frac 43}\Biggl[\biggl({\frac{1+x^2%
}{1-x}}\biggr)\biggl(\ln {\frac{\mu ^2}{m_c^2}}-1-2\ln (1-x)\biggr)\Biggr]%
_{+}\ .  \label{eq:fcc1}
\end{equation}
\end{Simlis}
Substituting Eqs.~(\ref{eq:Hardxsc0})-(\ref{eq:Hardxsc2}) into Eq.~(\ref
{eq:HadXsec4fl0}), the right-hand side can be re-organized as 
\begin{equation}
\begin{array}{lll}
F^{(4)}_{A,\lambda }(Q^{2},x,m_c,\mu) & = & \;\;f_A^g\otimes \,^1\omega
_{g,\lambda }^{c\bar{c}} \\ 
\; &  & +\,(f_A^c-\;f_A^g\otimes \,^1\tilde{f}_g^c-f_A^c\otimes \,^1\tilde{f}%
_c^c)\,\otimes \,^0\omega _{c,\lambda }^c \\ 
&  & +\;f_A^c\otimes \,^1\omega _{c,\lambda }^{cX} \\ 
&  & +\;\text{light-quark\ terms}\ ,
\end{array}
\label{eq:HadXsec4fl1}
\end{equation}
where the $\ln \left( \mu/ {m_c}\right) $ terms in the $\omega _{a,\lambda }$
factors are kept intact, and the needed subtraction terms are explicitly
grouped with the leading 2$\rightarrow $1 term with the same kinematics.
This is the form we use for the actual numerical calculations, which we
implement using a Monte Carlo approach.

It is useful to compare this calculation with the NLO 3-flavor calculation 
\cite{SvNorg} which can be written, in the same notation, as follows 
\begin{equation}
\begin{array}{lll}
F_{A,\lambda }^{(3)}(Q^{2},x,m_{c},\mu ) & = & \;\;f_{A}^{g}\otimes
\,^{1}\omega _{g,\lambda }^{c\bar{c}} \\ 
&  & +\;f_{A}^{g}\otimes \,^{2}\omega _{g,\lambda }^{gc\bar{c}} \\ 
&  & +\;f_{A}^{q}\otimes \,^{2}\omega _{q,\lambda }^{qc\bar{c}} \\ 
&  & +\;\text{light-quark\ terms}
\end{array}
\ .  \label{eq:HadXsec3fl}
\end{equation}
The term common to the two schemes is $\mathcal{O}\left( \alpha
_{s}^{1}\right) $ $\gamma ^{\ast }g\rightarrow c\bar{c}$, which appears as
the first line in both Eq.~\ref{eq:HadXsec4fl1} and Eq.~\ref{eq:HadXsec3fl}.
For this comparison, one can consider the rest of the terms in these
equations as complementary ``corrections'' to the first term. In particular,
the last two terms in the 3-flavor formula Eq.~\ref{eq:HadXsec3fl} are
genuine $\mathcal{O}\left( \alpha _{s}^{2}\right) $ corrections to the
common term in the threshold region; hence are commonly referred to as NLO.
But at high $Q^{2}\gg m_{c}^{2}$, these terms contain large logarithm
factors $\ln {Q^{2}}/{m_{c}^{2}}$ which vitiates the perturbation expansion.
They are \emph{not} NLO in this region.

In the 4-flavor calculation as organized in the form Eq.~\ref{eq:HadXsec4fl1}%
, the last two lines are also effectively $\mathcal{O}\left(
\alpha_{s}^{2}\right)$. This is because the distribution $f_{A}^{c}$ is
effectively $\mathcal{O}\left( \alpha _{s}^{1}\right)$ near threshold, and
there is a built-in cancellation between the leading order $\gamma ^{\ast
}c\rightarrow c$ partonic cross-section and the first subtraction term in
Eq.~\ref{eq:HadXsec4fl1}, as discussed in detail in Ref.\cite{ACOT}.
Although these $\mathcal{O}\left( \alpha_{s}^{2}\right)$ ``correction''
terms to the $\mathcal{O}\left( \alpha_{s}^{1}\right)$ common term (first
line) do not contain the full NLO $\mathcal{O}\left( \alpha _{s}^{2}\right)$
corrections at threshold, they do contain all the $\mathcal{O}\left(
\alpha_{s}^{2}\right)$ contributions which are enhanced by $\ln \left( {Q^{2}%
}/{m_{c}^{2}}\right)$ and quickly dominate as $Q^{2}$ increases. In fact,
these large logarithmic terms have been resummed to all orders in
perturbation theory via the DGLAP-evolved charm parton distribution
contributions in the second and third lines. Therefore, in the large $Q^{2}$
region the 4-flavor result represents a true NLO calculation.

In principle, if carried out to all orders in $\alpha _{s}$, the 3-flavor
and 4-flavor calculations in the form of $F_{A,\lambda
}^{(i)}(Q^{2},x,m_{c},\mu )$, $i=3,4$, would give exactly the same
prediction for all values of the arguments. The dependence on the scheme $%
(i) $ and the scale $(\mu )$ arises from the truncation of the perturbation
series. Therefore, in order to produce a physical prediction $F_{A,\lambda
}^{phys}(Q^{2},x,m_{c})$ and to relate the predictions in the two schemes, a
number of additional steps must be taken. In the presence of a non-zero
heavy quark mass, some of these steps are non-obvious; hence they can be the
source of confusion. An explicit discussion of these elements of the
calculation will make clear the flexibility as well as the uncertainties
inherent in the formalism. This we do in the Appendix, as part of the more
precise description of the formalism. Here, we mention only two features
which are particularly relevant for the subsequent discussions of physical
predictions.

First, within each scheme ($i=3$ or $4$), one needs to specify $\mu $ as a
function of the physical variables in order to make a \emph{physical
prediction}, i.e.\ 
\[
F_{A,\lambda }^{phys}(Q^{2},x,m_{c})=F_{A,\lambda }^{(i)}(Q^{2},x,m_{c},\mu
(x,Q,m_{c}))
\]
Although there is considerable freedom in choosing $\mu (x,Q,m_{c})$, two
conditions should be met so that the prediction can be reliable: (i) $\mu $
must be of the order of $Q$ or $m_{c}$ so that PQCD applies, and (ii) $%
F_{A,\lambda }^{(i)}(Q^{2},x,m_{c},\mu )$ must be relatively stable with
respect to variations of $\mu $ for the ($x,Q$)-range of interest. This is
the well-known \emph{scale-dependence} of any PQCD calculation. For the
problem at hand, a common choice for $\mu (x,Q,m_{c})$ is $\sqrt{%
Q^{2}+m_{c}^{2}}$: it represents the typical virtuality of the internal
parton lines in the important subprocesses. The presence of the uncertainty
associated with the choice of $\mu (x,Q,m_{c})$ in each scheme is
illustrated in Fig.~\ref{fig:cartoon} by the respective bands.\footnote{%
In the original ACOT paper, a different scale function $\mu (x,Q,m_{c})$ was
chosen: it was designed to enforce the condition $%
F^{(4)}(Q^{2},x,m_{c},\mu (x,Q,m_{c}))\rightarrow F_{LO}^{(3)}$ $%
(Q^{2},x,m_{c},m_{c})$ as $Q\rightarrow m_{c}.$ In retrospect, this choice
is artificial and unnecessary. The general formalism automatically ensures
that, $F^{(4)}(Q^{2},x,m_{c},\mu )\rightarrow F_{LO}^{(3)}$ $%
(Q^{2},x,m_{c},m_{c})$ as $\mu \rightarrow m_{c}$ for given $(Q^{2},x)$
to the order which we are working; but
it does not place any restriction on the behavior of $F^{phys}(Q^{2},x,m_{c})
$ as $Q\rightarrow m_{c}$ in any given scheme. As shown in Sec.\ \ref
{sec:inclusive}, the more natural choice of scale $\mu (x,Q,m_{c})=\sqrt{%
Q^{2}+m_{c}^{2}}$ leads to much improved physical predictions. \label%
{fn:AcotScale}}

Secondly, as shown in Fig.~\ref{fig:cartoon}, one needs to identify an
appropriate scale at which the 3-flavor and the 4-flavor scheme predictions
are both reasonable and mutually comparable, so that the transition from one
to the other in the composite scheme can be made smoothly. In the next
section we show that, for the inclusive charm production cross-section,
these conditions can be met over a rather large range of $Q$, extending down
to near the threshold region. One possibility then is to choose the
transition point (cf.\ Fig.~\ref{fig:cartoon}) at a low value, close to $m_c$%
, so that in effect the 4-flavor calculation by itself covers the full range
of physical interest.

In existing literature, it is already known that the 3-flavor order $%
\alpha_s^2$ (NLO) calculation can be extended to most of the currently
accessible energy scales without manifest ill-effects of the large
logarithms; and it agrees with data rather well. Its efficacy at very large $%
Q$ is not fully tested (cf.\ \cite{BMSvN} and results of next section). Our
calculation will demonstrate the robustness of the complementary, much
simpler (order $\alpha_s$) 4-flavor calculation. It is worthwhile pointing
out that, in the 4-flavor scheme, the order $\alpha_s$ calculation is, in
fact, also \emph{NLO} -- since the LO $\gamma^* c \rightarrow c$ term is of
order $\alpha_s^0$, as is the case in the standard QCD theory of inclusive
structure functions. This important point is discussed in detail in the
Appendix, Sec.\ \ref{sec:LoNlo}.

                      


\section{Inclusive Charm Structure Function}

\label{sec:inclusive}

In the previous section, the theoretical formulation of charm production in
DIS is presented within the context of the totally inclusive structure
functions $F_{\lambda }(x,Q)$. Although terms involving at least one charm
quark in the final state are the main subject of discussion, they have been
added to ``light quark contributions'' to form the totally inclusive
structure functions, cf.\ Eqs.~\ref{eq:HadXsec4fl1} and \ref{eq:HadXsec3fl}.
\ The reason to do this (rather than simply talk about an ``inclusive charm
structure function'', say $F_{2}^{c})$ is: charm quarks are not physically
observable; there is no unique or obvious definition of a ``$F_{2}^{c}$''
either experimentally or theoretically. Any experimental definition
necessarily depends on the procedure or prescription for tagging final state
charm (analogous to the jet-algorithm for defining jets). Any theoretical
definition will be scheme-dependent, and will be subjected to questions such
as infra-red safety (IRS), e.g.\ free from $\ln (m_{c})$ terms. The proper
matching of the experimental and theoretical definitions is also necessary
for a meaningful comparison of theory with experiment.

In this section, we shall be concerned with the comparison of our
calculation of inclusive charm production structure function with previous
calculations, such as that of ACOT \cite{ACOT} and that of the NLO
three-flavor scheme \cite{SvNorg}, as well as with existing experimental
results. Since the energy range where these comparisons will be made is
moderate, the question of infra-red safety is not a critical one -- as
evidenced by the general phenomenological success of the NLO 3-flavor
calculation (which is not IRS in the sense used in this paper). For this
purpose, it suffices to define the theoretical 4-flavor $F_{2}^{c}$ as the
quantity consisting of terms with at least one charm quark in the final
state in Eq.~\ref{eq:HadXsec4fl1}, i.e.\ it is the right-hand side less the
light-quark contribution. At the order we are calculating, this $F_{2}^{c}$
is actually well-defined and infra-red safe, -- there are no large logarithm
terms.\footnote{%
However, at the next order ($\alpha _{s}^{2}$) and beyond, $F_{2}^{c}$ is,
strictly speaking, not totally infra-red safe by itself (in the sense that
it does contain some un-cancelled $\log (\frac{Q}{m_{c}})$ factors); only
the sum with the light-quark contributions (in Eqs.\ref{eq:HadXsec4fl0},\ref
{eq:HadXsec4fl1}) is free from such potentially large logarithms. Thus, the
4-flavor formula without the light-parton term should not be used far beyond
the physical range $Q\sim \mu >m_{c}$ where it is well-defined. This issue
is discussed in Ref.~\cite{CSvN}, where a cutoff parameter is used to remove
the ``soft'' parts of the charm contribution from $F_2^c$ and to combine
them with the light-parton terms.} In addition, since the charm mass is just
about at the scale for PQCD to become valid and, as discussed at the end of
the previous section, the transition scale can in practice be chosen close
to $m_{c}$, \emph{the generalized \msbar\ calculation reduces in practice to
a 4-flavor (general mass) calculation}.

\paragraph{The 4-flavor NLO calculation:}

The parton distribution functions $f_{A}^{a}(x,\mu )$ needed for this
calculation must be defined in the same renormalization scheme as the hard
cross-sections used. In the 4-flavor general mass calculation, most CTEQ
parton distributions have been defined in the ACOT scheme as described in
the Appendix, with the matching scales chosen
at the heavy quark masses. In particular, the CTEQ5HQ set is generated with
non-zero heavy quark masses in the hard cross-sections, hence it is the
relevant one for our application. 
This set is evolved using NLO evolution,
which is appropriate to our calculation which is NLO at high energies.
For the numerical results presented below,
we use an updated version of CTEQ5HQ \cite{CTEQ5}. In order to compare with
results from the LO and NLO 3-flavor scheme, we need the corresponding
parton distributions in the 3-flavor scheme. The CTEQ5F3 distributions
satisfy this need, since they are obtained from global analysis of the same
data sets as the CTEQ5HQ set, but with parton distributions and hard
cross-sections in the 3-flavor scheme.

The calculation of the hard cross-section at $\mathcal{O}(\alpha_s)$ is
based on Eqs.~\ref{eq:Hardxsc1} and \ref{eq:Hardxsc2}. In the expression for
the overall structure functions, the combination $f_{A}^{g}\otimes
\,^{1}\omega _{g,\lambda }^{c\bar{c}}$ $+\,(f_{A}^{c}-\;f_{A}^{g}\otimes
\,^{1}\tilde{f}_{g}^{c}$) $\otimes \,^{0}\omega _{c,\lambda }^{c},$ due to
the subprocesses $\gamma ^{\ast }c\rightarrow c$ and $\gamma ^{\ast
}g\rightarrow c\bar{c},$ comprise the original ACOT calculation \cite{ACOT}.
With non-zero $m_{c},$ they are all finite. The implementation of these
terms in the new Monte Carlo calculation is straightforward. We have
verified that the new MC program reproduces the original ACOT results in
detail. Illustrations of the relative contributions of the various terms can
be found in Ref.\ \cite{ACOT}. Also, the smooth transition of the results of
the general formalism from the 3-flavor one near threshold to the 4-flavor
one at high energies is described in that paper.

The additional contributions $f_{A}^{c}\otimes \,^{1}\omega _{c,\lambda
}^{cX}-f_{A}^{c}\otimes \,^{1}\tilde{f}_{c}^{c}\otimes \,^{0}\omega
_{c,\lambda }^{c},$ due to the $\gamma ^{\ast }c\rightarrow gc$ subprocess
and virtual correction to the Born $\gamma ^{\ast }c\rightarrow c$ term, are
new for this calculation. They contain soft divergences which must be
cancelled. In this Monte Carlo implementation, we use the phase-space
splicing method to achieve the proper cancellation of the soft divergences
between the real and virtual parts. Details of this calculation are
contained in \cite{Xiaoning}. For double-checking, we have independently
implemented an analytic calculation based on the formulas of Hoffmann and
Moore \cite{HoffMoor}. The two calculations agree quite well with each other
over the full phase space, with the exception of small values of $Q/m_{c}$.
The difference could be attributed to a different treatment of the massive
charm quark kinematics adopted by Ref.~\cite{HoffMoor} in deriving their
formulas. This effect goes away when $m_{c}$ is small compared to $Q,$ as
expected.\footnote{%
The authors of Ref.\cite{KretzerS} have also identified some differences
between their recent calculation with Ref.\cite{HoffMoor}. We thank S.
Kretzer for providing us with some details of this comparison. Since the
difference in question is numerically not significant, none of the results
presented below are affected by this problem.}

\paragraph{Results and Comparison with order $\protect\alpha _{s}^{2}$
3-flavor calculation:}

We now present some typical numerical results on the theoretical $%
F_{2}^{c}(x,Q)$ obtained in our order $\alpha _{s}$ 4-flavor calculation
compared to those of order $\alpha _{s}^{2}$ 3-flavor scheme. The 3-flavor
results were obtained from the parametrization of Ref.~\cite{RiemS}.%
\footnote{%
We thank Brian Harris for furnishing us with his interface to this program.}
Each calculation corresponds to a different way of organizing the
perturbation series, hence has its natural region of applicability, as
discussed in the previous sections. To do a meaningful comparison, it is
important to take into account the estimated uncertainties of each
calculation. Following the example of Fig.~\ref{fig:cartoon}, we plot a band
for each of the predictions, obtained with two common choices of the
renormalization and factorization scales: $\mu =c\sqrt{Q^{2}+m_{c}^{2}}$
with $c=0.5$ and $2.0$. (This choice is more natural than that of the
original ACOT paper, c.f.\ footnote \ref{fn:AcotScale}.) The 3-flavor
calculation requires parton distributions defined in the same scheme. We use
the CTEQ5F3 set. The results vary in appearance, depending on the kinematic
variables. Two representative plots relevant for the HERA measurements are
shown in Fig.~\ref{fig:F2cTh} where $F_{2}^{c}(x,Q)$ is plotted against $Q$
for two values of $x$: (a) $x=0.01;$ and (b) $x=0.0001.$ These constitute
real examples of the cartoonistic Fig.~\ref{fig:cartoon}, which is designed
to emphasize the underlying ideas.\figFcTh These plots show that the
overlapping region of the two schemes is quite wide for $x=10^{-4}$; and the
overall behavior conforms with expectations. For $x=10^{-2}$, the
overlapping region is more limited and it is confined to low $Q$ values. The
3-flavor results fall below the 4-flavor ones at large values of $Q$ where
the latter should be more reliable. Compared to current data, both are
within the experimental range; the 4-flavor results are closer to the data
points, cf.\ next subsection, although the uncertainties of the 4-flavor
calculation are fairly large for $x=10^{-4}$.

Since the order $\alpha_s$ 4-flavor results are comparable to the order $%
\alpha_s^2$ ones at energy scales close to the threshold in both cases (and
for other values of $x$), it is reasonable to choose the transition scale at
a relatively low value, as mentioned earlier in the paper. The band
representing the 3-flavor calculation does become wider at large Q for $%
x=10^{-2}$ (where the absolute values also are lower than the 4-flavor
calculation and data); but for $x=10^{-4}$, it remains quite narrow. Thus,
the theoretically infra-red unsafe logarithms, $\ln^{1,2}(\mu/m_c)$, do not
seem to cause serious problems, at least for very low $x$.

\paragraph{Results Compared to recent Zeus data:}

The general agreement between existing data on charm production with order $%
\alpha_s^2$ 3-flavor calculations, using the scale choice $\mu = \sqrt{%
Q^{2}+m_{c}^{2}}$, is well known. It is of interest to compare the same data
on ``inclusive charm production structure function'' $F_{2}^{c}$ with our
order $\alpha_s$ 4-flavor calculation. Fig.~\ref{fig:IncExpTh} compares the
results of our calculation, using the same scale choice, with data from the
recent ZEUS \cite{ZEUScharm97} data. The agreement is clearly excellent.
Data also agree with the order $\alpha_s^2$ 3-flavor calculation as shown in 
\cite{ZEUScharm97}. This higher order calculation is obviously much more
elaborate than the order $\alpha_s$ 4-flavor calculation presented here. We
see that, for inclusive cross-sections, the resummation of the $\alpha_s^n
ln^n(\mu/m_c)$ terms into the charm distribution function $f^c(x,\mu)$ in
the 4-flavor scheme offers a more efficient way to organize the perturbative
series, resulting in an effective NLO calculation already at order $\alpha_s$%
, cf.\ Sec.\ \ref{sec:LoNlo}.\figIncExpTh




\section{Semi-inclusive Cross Sections with Tagged Charm Hadrons}

\label{sec:tagged}

\subsection{General considerations}

Now we consider semi-inclusive cross sections, with a charm hadron tagged in
the final state. Naively, to compute this cross section, one simply convolves
the cross sections for parton final states, Eqs.~\ref {eq:HadXsec4fl1} and
\ref{eq:HadXsec3fl}, with a suitable fragmentation function of partons into the
final charm hadron. However, the factorization of the final state particles
through fragmentation functions is only rigorously defined in the limit
$Q^{2}\gg m_{c}^{2}$. Thus, the treatment of tagged charm particles in the
final state can only be systematically applied at high energies, using the
4-flavor scheme. However, it is a common practice to introduce fragmentation
functions into charm hadrons even in the 3-flavor scheme, and for energies not
far above threshold. This approach should be considered a convenient
phenomenological model of hadronization, perhaps adequate for current
experimental accuracy, rather than rigorous theory. We follow this practice in
our calculation, employing fragmentation functions over the full range of
$Q^{2}$, while maintaining the correct factorization-scheme implementation at
high $Q^{2}$.

The fragmentation functions $d_{a}^{H}(x,\mu )$ obey the standard
(mass-independent) QCD evolution equations, and are determined from suitable
initial functions at some given scale $\mu= Q_{0}$, of the order of $m_{c}$.
Following Mele and Nason \cite{MelNas}, for a given final-state charm hadron
${H},$ we write
\begin{equation}
d_{a}^{H}(x,Q_0)=d_{a}^{c}(x,Q_0)\otimes D_{c}^{H}(x,Q_0) \label{eq:HadFrag}
\end{equation}
where the partonic charm fragmentation functions $\left\{
d_{a}^{c};a=l,c\right\} $ are perturbatively calculable, and $%
D_{c}^{H}(x,Q_0)$ is considered non-perturbative and is to be obtained by
comparison with experiment.

For the perturbatively calculable fragmentation functions, Ref.~\cite{MelNas}
gives, to order $\alpha _{s}$:\
\begin{eqnarray}
d_{c}^{c}(x,Q_0) &=&\delta (1-x)+{\frac{{\alpha _{s}(Q_0)C_{F}}}{{%
2\pi }}}\left[ {\frac{{1+x^{2}}}{{1-x}}}\left( \ln {\frac{{Q_0^{2}}}{{%
m_{c}^{2}}}}-2\ln (1-x)-1\right) \right] _{+}  \nonumber \\
d_{g}^{c}(x,Q_0) &=&{\frac{{\alpha _{s}(Q_0)T_{F}}}{{2\pi }}}%
(x^{2}+(1-x)^{2})\ln {\frac{{Q_0^{2}}}{{m_{c}^{2}}}} \label{eq:pertFrag}
\\ d_{q,\bar{q},\bar{c}}^{c}(x,Q_0) &=&0  \nonumber
\end{eqnarray}
where $T_{F}=1/2$ and $C_{F}=4/3$. \ In keeping with the choice of the matching
scale in our overall calculation, we choose $Q_0=m_{c}$ for convenience in this
paper.

For the non-perturbative charm quark into charmed mesons fragmentation
function $D_{c}^{H}(z),$ we used the conventional Peterson form \cite
{Peterson},
\begin{equation}
D_{c}^{H}(z)=\frac{A}{z[1-1/z-\epsilon /(1-z)]^{2}},  \label{eq:peterson}
\end{equation}
For the charm meson $D^{(\ast )},$ which will be our focus because of available
experimental data, we take $\epsilon =0.02,$ cf.\ \cite{Cacciari:1997ad}, and a
value for $A$ such that the branching fraction $B(c\rightarrow D^{\ast })=0.22$
\cite{OPAL95}. We note that, although the perturbative fragmentation functions,
Eq.\ \ref {eq:pertFrag}, contain singular (generalized) functions, the overall
parton-to-charm-meson fragmentation functions $d_{a}^{H}(x,\mu),$ Eq.\
\ref{eq:HadFrag}, are well behaved after convolution with the above
non-perturbative fragmentation function $D_{c}^{D^{(\ast )}}(z)$.

In principle, after evolving to high enough $Q^{2}$ so that $\alpha _{s}\log
(Q^{2}/m_{c}^{2})$ is of order one, all of the fragmentation functions $%
d_{c}^{H},\ d_{g}^{H},\ d_{q,\bar{q},\bar{c}}^{H}$ eventually become of the
same size. \ In practice, however, at HERA energies we find $d_{c}^{H}\gg
d_{g}^{H}\gg d_{q,\bar{q},\bar{c}}^{H}$. For currently required accuracy, it
suffices to keep only the charm-to-hadron contributions, proportional to $%
d_{c}^{H}$. In this approximation, our calculation of the cross section with
a tagged charm hadron can be written in the 4-flavor scheme as
\begin{equation}
\begin{array}{lll}
F_{A,\lambda }^{H}(Q^{2},x,..) & = & \;\;f_{A}^{g}\otimes \,^{1}\omega
_{g,\lambda }^{c\bar{c}}\otimes \,d_{c}^{H} \\
\; &  & +\,(f_{A}^{c}-\;f_{A}^{g}\otimes \,^{1}\tilde{f}_{g}^{c}-f_{A}^{c}%
\otimes \,^{1}\tilde{f}_{c}^{c})\,\otimes \,^{0}\omega _{c,\lambda
}^{c}\otimes \,d_{c}^{H} \\
&  & +\;f_{A}^{c}\otimes \,^{1}\omega _{c,\lambda }^{cX}\otimes \,d_{c}^{H}\
.
\end{array}
\label{eq:HadXsec4fl1Hc}
\end{equation}

To ensure that this calculation is adequate, we have also calculated the
contribution from one of the more important remaining subprocesses: gluon
fragmentation in an order $\alpha _{s}$ light parton hard scattering, i.e.\ $%
\gamma ^{\ast }q\rightarrow gq\ ;\ g\rightarrow H$. It is given by: $%
f_{A}^{q}\otimes \,^{1}\widehat{\omega }_{q,\lambda }^{qg}\otimes \,d_{g}^{H}
$, where $d_{g}^{H}$ is the gluon fragmentation function computed from
Eqs.~\ref{eq:HadFrag} and \ref{eq:pertFrag}. We have verified that its
contribution remains small throughout the current energy range. It becomes more
noticeable only in the large $Q$ limit. However, in this limit, the gluon
fragmentation function term is not infra-red safe by itself. To insure
consistency at high energies, one needs to include a full set of infra-red safe
higher order subprocesses along with it. The full calculation is more
appropriately considered as part of the next order project.

At the same level of accuracy, it is also reasonable to ignore the evolution of
$d_{a}^{H}(x,\mu(x,Q))$, since the effect of QCD evolution is not significant
over the currently accessible HERA $Q$ range. One can use the un-evolved
$d_{c}^{H}(x,\mu=Q_0)$ in place of the fully evolved $d_{c}^{H}(x,\mu(x,Q,m_c))
$ with much gain in efficiency of calculation and little sacrifice in accuracy.
The error incurred is of the same order as that incurred by
neglecting the subleading fragmentation functions, $d_{g}^{H}$ and $d_{q,%
\bar{q},\bar{c}}^{H}$ ; and the comments on accounting for $\ln (m_{c})$
factors made there also apply here. To be sure about this, we have performed
the calculation both with and without evolving $d_{c}^{H}.$  The difference is
indeed small. Therefore, in all our subsequent plots we shall include only the
direct charm-to-hadron contributions of Eq.~(\ref{eq:HadXsec4fl1Hc}) with the
un-evolved fragmentation function $d_{c}^{H}(x,Q_{0})$.


\subsection{Differential distributions}

We employ the Monte Carlo method to carry out the numerical phase-space
integration; the new program package is implemented in the C++ programming
language. Therefore, we can generate differential distributions involving
final-state charm mesons, incorporating kinematic cuts appropriate for
specific experimental measurements, in addition to fully inclusive
cross-sections.

In working with on-mass-shell heavy flavor quarks and hadrons in the parton
language, there is an ambiguity in defining the momentum fraction variables $%
x\;(z)$ for the parton distribution (fragmentation) function. \ This problem
arises in all schemes; and it goes away at high energies ($Q\gg m_{c}$), where
the parton picture becomes accurate.\ Following the modern practice in proofs
of factorization, we define the momentum fraction variables as ratios of the
relevant light-cone momentum components, e.g.$\;p_{D}^{+}=zp_{c}^{+}$ for
fragmentation of a charm quark into a $D$ meson. Other authors, e.g.\
\cite{FMNR97,HarrSmith}, use the prescription $\vec{p}_{D}=z\vec{p}_{c}$ and
adjust the energy variable to enforce the mass-shell condition. At moderate
energies, any noticeable differences in results due to the choice of this
prescription signals that the calculation using fragmentation functions is
outside the region of applicability of the parton formalism. We have verified
that the results presented below are insensitive to the choice
between the two prescriptions.\footnote{%
There are some differential distributions, especially those associated with
the unphysical partons (such as momentum fraction carried by the charm
quark, sometimes seen in the literature) which are more sensitive to the
choice of definition of the momentum fraction variable.}

The QCD formula, Eq.\ \ref{eq:HadXsec4fl1Hc}, contains three scale choices in
principle: the renormalization scale, the factorization scale and the
fragmentation scale. For simplicity, we choose the same energy scale
$\mu(x,Q,m_c)$ for all three. As in the case of the inclusive $F_2^c(x,Q)$, for
results shown below, we choose the simple function $\mu=c (Q^2+m_c^2)^{1/2}$,
which characterizes the typical virtuality of the process. The constant $c$ is
of order 1; and is varied over same range when we try to estimate the
scale-dependence of the physical predictions. The magnitude of the charm
cross-section is sensitive to the value of $m_c$. For results presented here,
we use $m_c^{\overline{\rm MS}}=1.3 GeV$, the value used in the CTEQ5HQ parton
distribution analysis (which is in the middle of the range given by the PDG
review).

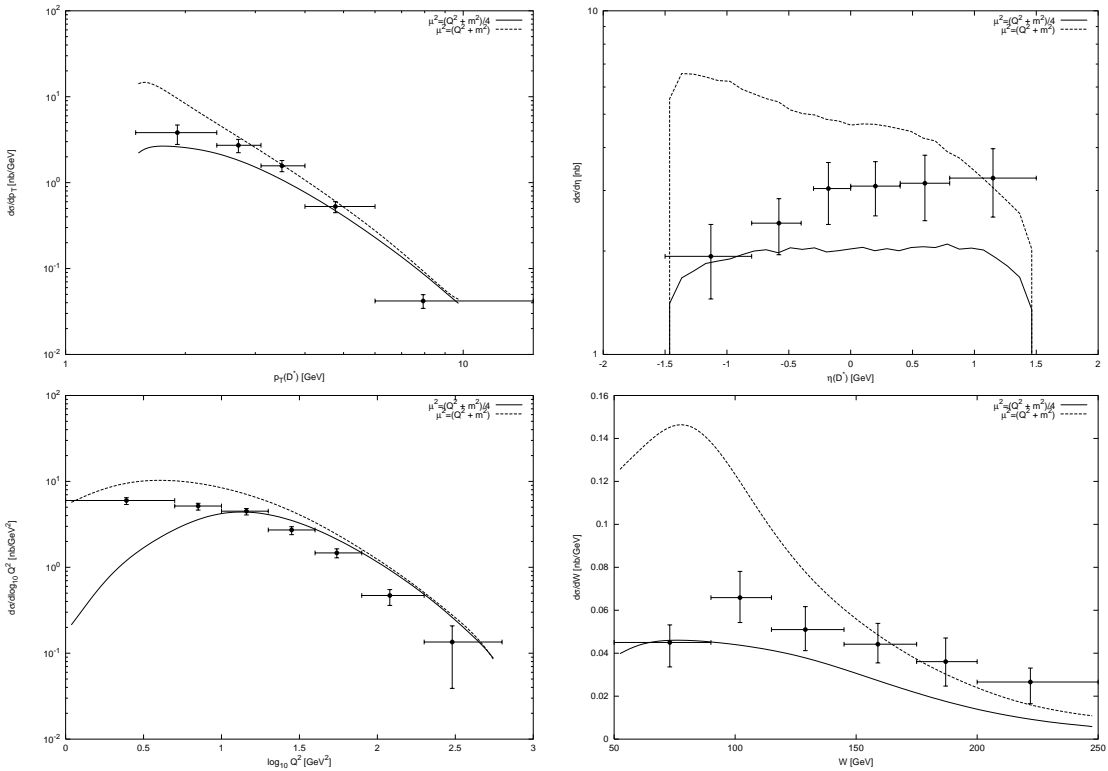
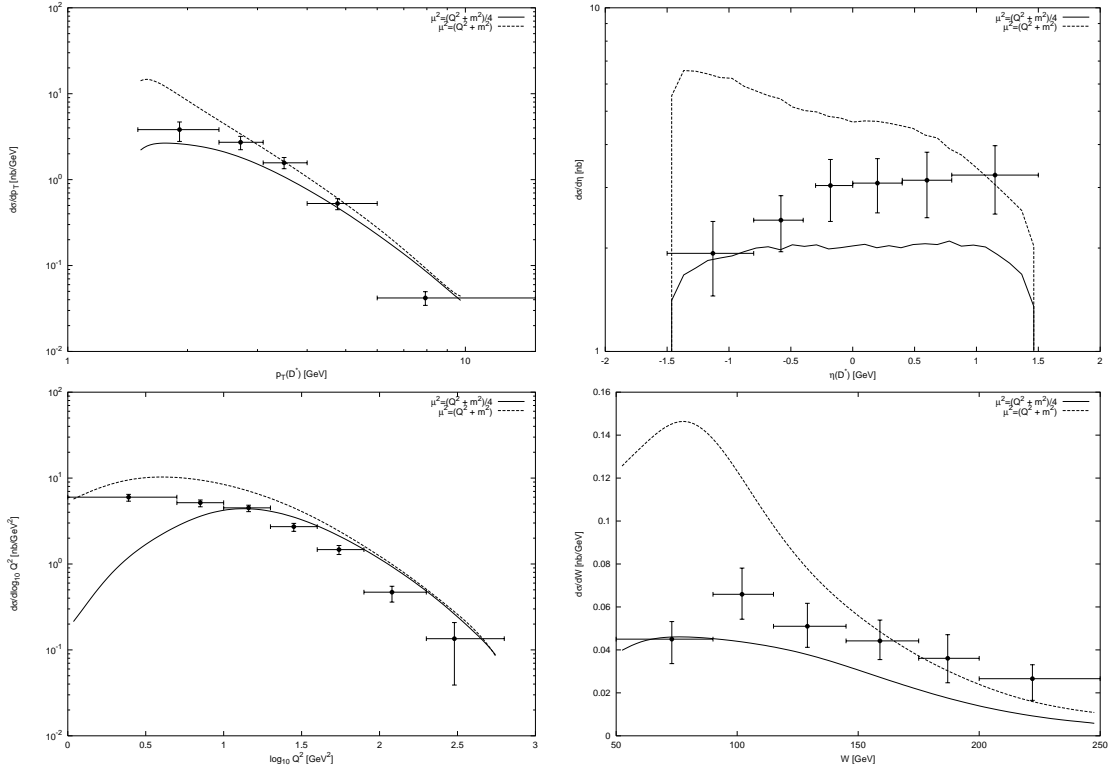
Fig.\ \ref{fig:DiffDis} shows plots of four differential distributions for $%
D^{\ast }$ production at HERA, calculated using the NLO ($\alpha _{s}$)
generalized \msbar\ 4-flavor formalism described above. \ The kinematic
variables and their ranges correspond to those of the 1996-97 ZEUS
data: $1<Q^{2}<600\;$GeV$^{2}\;;\;0.02<y<0.7\;;\;1.5<p_{T}^{D^{\ast }}<15\;$%
GeV ; $\left| \eta _{D^{\ast }}\right| <1.5.$ Each distribution contains two
curves obtained with two values of the constant $c=0.5,1$ in the definition of the
scale parameter described above. These predictions, using the CTEQ5HQ parton
distributions, are compared to the ZEUS data \cite {ZEUScharm97}. We observe a
rather large scale dependence in these results. This is not surprising, since
the compensation among the various subprocesses which underlie the
scale-independence of the physics predictions (up to some order of perturbation
theory), strictly speaking, only apply to the inclusive cross-section. The
experimental kinematical cuts implemented in these exclusive calculations to
some extent undermine the mutual cancellation between diagrams which are
necessary for relatively scale-independent predictions. For example, the order
$\alpha_s^0$
$\gamma^* c \rightarrow c$ HE term (which resums the logarithms arising from
the near-collinear configurations of an infinite tower of higher-order
diagrams) implements the full contribution in collinear kinematics, a clear
over-simplification.

Keeping this fact in mind, and with current relatively large experimental
errors, the results of Fig.\ \ref{fig:DiffDis} can be considered rather
encouraging: the $Q^2$ and $p_T$ distributions show very good general
agreement; while the $W$ and $\eta_D$ distributions are ``in the right ball
park'', the shapes are too scale-dependent to allow for meaningful
``predictions''. (A specific choice of scale, in between the two shown, will
actually yield theory curves in reasonable agreement with data, within errors.)
In order to make genuine predictions on differential distributions in the
4-flavor scheme, it is necessary to extend the calculation to order
$\alpha_s^2$, which would be NNLO in the 4-flavor scheme.  This can be done by
transforming already available NLO results for 3-flavor calculations into the
4-flavor scheme. At the same order in $\alpha_s$, the 4-flavor scheme
calculation is, of course, more involved than the (NLO) 3-flavor one because of
the need for including the necessary subtraction terms in a NNLO calculation --
such as those appearing in Eqs.\ \ref{eq:Hardxsc1}, \ref{eq:Hardxsc2}, and
\ref{eq:HadXsec4fl1}.

The calculation of these differential distributions in the $\alpha _{s}^{2}$
3-flavor FFN scheme was carried out by \cite{HarrSmith}. Generally good
agreement between these calculations and the recent ZEUS data, using specific
parton distributions, scale choices, etc.\ has been reported in Ref.\ \cite
{ZEUScharm97}. Although the dependence of the predictions to the scale choice
was not discussed in this comparison, it is relatively mild, according to
\cite{HarrSmith}. This is to be expected, because the sensitivity to cuts is
reduced with a better approximation of the final state particle configurations
provided by the order $\alpha _{s}^{2}$ calculation. This fact implies greater
predictive power for the differential distributions than the order $\alpha_s$
4-flavor calculation.

The $\eta _{D^{\ast }}$ distribution in both the 3-flavor and the 4-flavor
calculations appear to differ in shape compared to the existing data points.
This could be due to the inadequacy of applying the fragmentation function
approach at less than asymptotic region, as discussed in the beginning of this
section. In particular, if the $D^{\ast }$ is not collinear to the parton, as
assumed in this approach, the rapidity distribution will be affected. More
extensive study of this effect is obviously needed.

\figDiffDis

\section{Conclusions}

\label{sec:conclude}

In this work we have shown how the generalized \msbar\ formalism can be used
to calculate the production of a heavy quark over a wide range of energies,
from threshold to high $Q^2$.  For charm production, the formalism consists of a 3-flavor scheme calculation at low energies, a
4-flavor scheme calculation at high energies, a matching condition between the
two schemes, and a transition scale chosen at which one switches between the
two schemes.  Specifically, we have extended the original
ACOT calculation for charm production at HERA by adding those terms which
are necessary to bring the 4-flavor part of the calculation to NLO accuracy
at high energies.  This brings our calculation to the same level of accuracy
as the other theoretical inputs to the CTEQ and MRS global QCD analyses.

The generalized \msbar\ formalism is ideally suited for inclusive calculations,
such as $F_2(x,Q)$ in DIS.  We have shown that for a physically-motivated choice of the renormalization/factorization scale, $\mu$, the transition scale can
be chosen rather low, so that our 4-flavor scheme can be used in the HERA energy range with excellent agreement compared to data, and with considerable economy compared to the more elaborate NLO 3-flavor calculation.  The calculation is
much simpler and the resulting program runs faster.  Furthermore, the NLO
corrections are much smaller in the 4-flavor calculation.  This suggests that
the resummation involved in the charm quark distribution function picks up
the most important higher-order corrections even at modest energies.  It also
indicates that the perturbation series in the 4-flavor calculation
is very well-behaved.

Our calculation has been implemented as a Monte Carlo program, so that we
have also calculated differential distributions for exclusive charm final
states.  By incorporating experimental cuts in the Monte Carlo we are
able to ensure that our calculation is compared directly with the data,
without the need for any theoretical extrapolation to all of phase space.
The agreement with HERA data is reasonable.  However, in this case the 3-flavor
NLO calculation has somewhat of an edge, because there is no approximation
on the kinematics as is necessary in the resummation used in the 4-flavor
calculation.

The comparison between available data and the order $\alpha_s$ 4-flavor and
order $\alpha_s^2$ 3-flavor calculations discussed in the last two sections
demonstrates the complementary nature of the two schemes -- both with regard to
the kinematic regions and to the physical quantities they are suitable for. \
With more abundant and more precise experimental data on charm production, the
composite generalized \msbar\ formalism, which encompasses both, will be needed to make reliable
comparisons.  With this in mind we are now ready to extend the 4-flavor
calculation to include the necessary ${\cal O}%
(\alpha_{s}^{2})$ terms, along with the ${\cal O}%
(\alpha_{s}^{2})$ matching conditions.
Such a calculation would include all the
advantages the current calculation has in addition to the advantages of the
current 3-flavor NLO calculation.

Finally, the generalized \msbar\ formalism provides a framework to extract
information on the gluon and the charm distributions of the nucleon -- to the same
accuracy as the overall NLO global QCD analysis based on total DIS structure
functions and other hard processes.
Having established that our calculation does a reasonable job in
describing the existing HERA data, we are now in a position to explore the
question of whether the proton contains a non-perturbative charm component
\cite{Brodsky:1981se}.
Certainly, it must at some level, so the real question is how small is it? The handle on the charm distribution
is unique to the generalized formalism, since the fixed 3-flavor scheme does
not allow the charm
parton as an independent degree of freedom.%
\footnote{%
The CTEQ5HQ parton distributions used in our calculations here contains charm
partons, but does not use an independent non-perturbative charm component as
input. This feature is not inherent to the formalism.} We note that there have
been recent phenomenological studies of ``intrinsic charm'' which take the
theoretical cross-section to be the simple sum of the 3-flavor FFN scheme
formulas and an intrinsic charm contribution by the heavy-quark excitation
mechanism \cite{HSV96,GV97,MT97}. This approach cannot be internally
consistent, because the 3-flavor calculation assumes parton evolution with no
charm quark distribution, while ``intrinsic charm'' explicitly requires one.
The intimate interplay between the hard matrix elements and parton evolution is
consistently incorporated only in the generalized \msbar\ scheme. Although
there has been a recent effort to examine this problem incorporating some of
the ideas of the general scheme \cite{Steffens}, a fully consistent
study, preferably based on more extensive data, still awaits to be done.%
\footnote{%
Because all active partons are coupled, the inclusion of a non-perturbative
charm component of the nucleon will affect all parton distributions.  Hence, it
will influence the global fitting of all available data sets, including the
extensive DIS sets.  If a reliable conclusion is to be drawn, it will not
suffice to combine a subset of existing parton distributions (in a conventional
scheme) with modified charm and gluon distributions (in the new scheme), as is
done in \cite{Steffens} using GRV and MRST distributions. Phenomenologically,
such a combination can upset the original good global fit to existing data,
particularly the precision DIS data.  Among theoretical problems, the NLO
evolution equation and the momentum sum rule will not be observed, due to the
mixing of these hybrid parton distributions.
}%

\subsection*{Acknowledgments}

We would like to thank Brian Harris for many helpful discussions and especially
for providing us with the code for the NLO 3-flavor calculations, and for
verification of the numerical results. We would also like to thank John Collins
and Fredrick Olness for many discussions and help; Jim Whitmore for assistance
about the ZEUS data; and Jack Smith for discussions on the extensive order
$\alpha_s^2$ calculations carried out by the Stony-Brook/Leiden group and for
clarification of the proper references to their work.
This work was supported in part by the US National Science Foundation under
grants PHY9802564 and PHY9722144.
%

\newpage
\input{lhk4.app}  

\input{lhk4.cit}   


\end{document}